# Half-Duplex and Full-Duplex AF and DF Relaying with Energy-Harvesting in Log-Normal Fading

Khaled M. Rabie, *Member, IEEE*, Bamidele Adebisi, *Senior Member, IEEE*, Mohamed-Slim Alouini, *Fellow, IEEE*

*Abstract*—Energy-harvesting (EH) and wireless power transfer in cooperative relaying networks have recently attracted a considerable amount of research attention. Most of the existing work on this topic however focuses on Rayleigh fading channels, which represent outdoor environments. In contrast, this paper is dedicated to analyze the performance of dual-hop relaying systems with EH over indoor channels characterized by log-normal fading. Both half-duplex (HD) and full-duplex (FD) relaying mechanisms are studied in this work with decode-and-forward (DF) and amplify-and-forward (AF) relaying protocols. In addition, three EH schemes are investigated, namely, time switching relaying, power splitting relaying and ideal relaying receiver which serves as a lower bound. The system performance is evaluated in terms of the ergodic outage probability for which we derive accurate analytical expressions. Monte Carlo simulations are provided throughout to validate the accuracy of our analysis. Results reveal that, in both HD and FD scenarios, AF relaying performs only slightly worse than DF relaying which can make the former a more efficient solution when the processing energy cost at the DF relay is taken into account. It is also shown that FD relaying systems can generally outperform HD relaying schemes as long as the loop-back interference in FD is relatively small. Furthermore, increasing the variance of the log-normal channel has shown to deteriorate the performance in all the relaying and EH protocols considered.

## I. INTRODUCTION

In conventional energy-constrained wireless networks, the network connectivity and operability is traditionally maintained from manually recharging or replacement of batteries which can be in many scenarios inconvenient or even impossible in others. Scavenging energy from the surrounding environment, also commonly known as energy harvesting (EH), using for instance solar power, thermal energy or wind energy seems to offer a promising and cost-effective solution that can prolong the life-time of energy-constrained wireless devices. However, such uncontrollable natural energy resources can not guarantee a constant amount of energy which may put reliable communications at risk. To overcome this limitation, EH from man-made radio-frequency (RF) signals has recently been proposed since such signals not only can carry information but can also serve as a convenient energy source; this is referred to as wireless power transfer.

Khaled M. Rabie and Bamidele Adebisi are with the school of Electrical Engineering, Manchester Metropolitan University, Manchester, M15 6BH, UK. (e-mails: k.rabie@mmu.ac.uk; b.adebisi@mmu.ac.uk,).
Mohamed-Slim Alouini is with King Abdullah University of Science and Technology (KAUST), Thuwal, Mekkah Province, Saudi Arabia. (e-mail: slim.alouini@kaust.edu.sa).

This concept has particularly generated considerable research interest in the so-called simultaneous wireless information and power transfer (SWIPT) networks. Although many studies have analyzed the performance of point-to-point SWIPT based systems [1]–[5], cooperative relaying SWIPT networks, where an intermediate relaying node is used to forward the source's information to the intended destination, have been by far more extensively investigated in the literature, see e.g., [6]–[8] and the reference therein. More specifically, the authors in [7] examined the performance of a half-duplex (HD) amplify-and-forward (AF) relaying network with EH where a greedy switching policy was deployed. Later on, the authors of [8] evaluated the performance of a one-way AF relaying system with three different EH protocols, namely, time-switching relaying (TSR), power-splitting relaying (PSR) and ideal relaying receiver (IRR). Furthermore, [9] considered the outage probability and ergodic capacity analysis of a two-way EH relay network. Decode-and-forward (DF) relaying with TSR and PSR EH was studied in [10] where the authors derived exact analytical expressions for the achievable throughput and ergodic capacity. Other works studying DF EH systems have also appeared in [11]–[14]. The performance of energy-constrained multiple-relay networks with relay selection is examined in [15]. Physical later security (PLS) HD EH-based multi-antenna AF relay networks was studied in [16]. In this work, the authors exploited the artificial noise signal, which is traditionally used to improve the secrecy rate, to assist in powering the energy-constrained relay. Similar works combining PLS and EH have been reported in [17]–[19].

The aforementioned relaying SWIPT systems have been limited to HD relaying costing 50% loss in spectral efficiency. Therefore, full-duplex (FD) relaying mechanism, which exploits the scarce frequency spectrum more efficiently by supporting simultaneous signal transmission and reception over the same frequency band [20]–[22], has recently been implemented in SWIPT networks, see e.g., [23]–[28].

All the existing works on SWIPT, including the ones above, have been limited to a restricted number of fading models such as Rayleigh, Nakagami-m and Rician, which are valid to model the outdoor wireless channel. On the other hand, the analysis of SWIPT systems over indoor log-normal fading channels is scarce; in fact, to the authors' best knowledge, only one study has recently appeared in the literature investigating the performance of a HD SWIPT network with AF relaying over the log-normal fading channel [29]. It is worth noting that log-normal fading is usually used to study the communication performance in many reference scenarios. For instance, it can accurately characterize shadowing from obstacles and moving human bodies in indoor environments, and log-normal distribution offers a better fit for modeling fading fluctuations in indoor wireless channels [30]–[34]; it is also used to model small-scale fading for indoor ultra-wideband (UWB) communications [35], [36]. Furthermore, empirical fading channel measurements have shown that short-term and long-term fading effects over the slowly-varying indoor channel tend to get mixed and log-normal statistics become dominate; hence, it describes the distribution of the channel path gain [33], [37], [38]. RF signals in indoor wireless channels may strongly attenuate due to obstacles such as object mobility and building walls which necessitates the use of relays [39], [40].

Motivated by the above considerations, this paper is therefore dedicated to study the performance of relaying SWIPT systems over indoor log-normal channels with HD and FD; both AF and DF relaying schemes are adopted along with TSR, PSR and IRR EH protocols. The system performance is evaluated in terms of the ergodic outage probability[1].

Therefore, the contribution of this paper is as follows. First, we derive analytical expressions of the ergodic outage probability for a dual-hop HD SWIPT system in log-normal fading with both DF and AF relaying. In this respect, TSR, PSR and IRR EH protocols are investigated for each case; hence, six distinct system configurations are studied resulting in six different analytical expressions. The second part of this work deals with FD SWIPT in log-normal fading for both DF and AF relaying. The other contribution resides in examining the impact of log-normal fading parameters on the ergodic outage probability as well as comparing the performance of DF and AF relaying in various EH protocols. Furthermore, the optimization problem of the EH time factor and power-splitting factor in the TSR and PSR schemes is addressed. Results show that when the processing energy cost of the DF relay is ignored, DF-based systems always offer better performance compared to that of AF relaying. It is also demonstrated that increasing the log-normal fading channel variance leads to performance degradation in all systems under study. In addition, the FD systems tend to outperform the HD ones given that the loop-back interference, caused by FD relaying, is relatively small.

The following notations are used in this paper. $f_X(\cdot)$, $F_X(\cdot)$ and $\bar{F}_X(\cdot)$ denote the probability density function (PDF), the cumulative distribution function (CDF) and the complementary CDF (CCDF) of the random variable (RV) $X$, respectively. $E\{\cdot\}$ and $\min\{\cdot\}$ denote the expectation operator and the minimum argument, respectively.

The rest of this paper is organized as follows. Section II briefly describes the HD and FD system models. Section III analyzes the ergodic outage probability performance of the dual-hop HD SWIPT network with DF and AF relaying, and TSR, PSR and IRR EH protocols. Section IV is dedicated to study FD with DF and AF relaying over log-normal fading channels. Numerical examples and simulation results are presented and discussed in Section V. Finally, conclusions are drawn in Section VI.

[1]Note that part of this paper was presented at the IEEE (GLOBECOM 2016) [41].

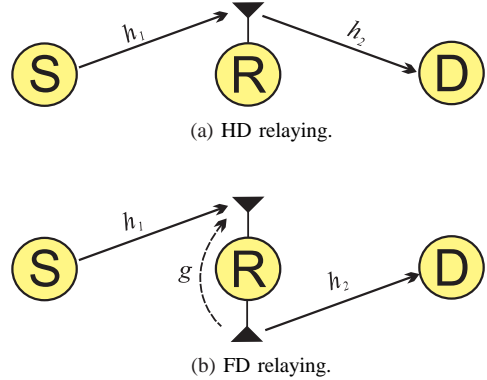

Figure 1: Basic bock diagram of the HD and FD relaying systems under consideration.

## II. SYSTEM MODEL

Fig. 1 illustrates the basic system diagrams of the considered dual-hop HD and FD systems which consist of a source node, relay node and destination node. The source first transmits its data, with power $P_s$, to the destination via an intermediate energy-constrained relay. The relay can be either based on DF or AF. It is assumed that there is no direct link between the end nodes and that the relay is powered entirely from harvesting the energy signal transmitted by the source node. The source-to-relay and relay-to-destination channel coefficients are denoted by $h_1$ and $h_2$ with $d_1$ and $d_2$ being the corresponding distances, respectively. Although the power consumed by the circuitry to process data at the relay is neglected in our derivations, this will be discussed in depth in the results section.

In the HD scenario, see Fig. 1(a), the relay has a single-antenna and hence the source-to-destination information transmission is accomplished over two phases. On the other hand, in FD relaying, Fig. 1(b), the relay is equipped with two antennas which allows simultaneous information reception and transmission at the relay with a loop-back interference channel denoted as $(g)$. It is important to mention that real channels are considered throughout. Note that in both HD and FD cases, the source and destination nodes are equipped with a single-antenna each; this configuration has been adopted in several studies dealing with log-normal fading channels [35], [42], [43]. In FD relaying, one antenna is dedicated to harvesting energy, and is only used for this purpose. This configuration is chosen not only for its relative ease of implementation but also because, according to [44], it attains comparable performance to the case when the two antennas are exploited for EH.

We assume that $h_1^2$ and $h_2^2$ are independent and identically distributed log-normal RVs with parameters $LN\left(2\mu_{h_1}, 4\sigma_{h_1}^2\right)$ and $LN\left(2\mu_{h_2}, 4\sigma_{h_2}^2\right)$, respectively, where $\mu_{h_i}$ and $\sigma_{h_i}$ (both in decibels) are respectively the mean and the standard deviation of $10\log_{10}(h_i)$, $i \in \{1, 2\}$. In addition, the loop-back interference channel $g^2$ is assumed log-normally distributed with parameters $LN\left(2\mu_g, 4\sigma_g^2\right)$; note that this is a key parameter determining the strength of the loop-back interference and hence the overall performance of FD relaying.

As mentioned in the introduction, the system performance is evaluated in terms of the ergodic outage probability. This probability is defined as the probability that the instantaneous capacity falls below a certain threshold value ($C_{th}$) and can be calculated for the AF and DF relaying systems, respectively, as

$$\mathcal{O}(C_{th}) = \Pr\{C_d(\gamma_d) < C_{th}\}, \quad (1)$$

and

$$\mathcal{O}(C_{th}) = \Pr\{\min\{C_r(\gamma_r), C_d(\gamma_d)\} < C_{th}\}, \quad (2)$$

where $C_r$ and $C_d$ are the instantaneous capacities at the relay and destination nodes, respectively, while $\gamma_r$ and $\gamma_d$ denote the corresponding signal-to-noise ratios (SNR).

The received signal at the relay during the EH phase in both HD and FD can be expressed as

$$y_r(t) = \sqrt{\frac{P_s}{d_1^m}} h_1 s(t) + n_a(t), \quad (3)$$

where $m$ is the path loss exponent, $s(t)$ is the information signal normalized as $\mathbb{E}\left[|s|^2\right] = 1$ and $n_a(t)$ is narrowband Gaussian noise introduced by the receiving antenna at the relay with variance $\sigma_a^2$. We next derive analytical expressions of the ergodic outage probability for the systems under study.

## III. HALF-DUPLEX RELAYING SYSTEM

This section analyzes the performance of HD relaying over log-normal channels with both DF and AF relaying.

### A. Half-Duplex with DF Relaying

Below, we derive analytical expressions for the HD-DF system with TSR, PSR and IRR EH protocols.

*1) HD-DF-TSR System :* In the TSR protocol, the time required to transmit one block from the source to the destination, also referred to as the time frame ($T$), is divided into three time slots as shown in Fig. 2. The first time period is the EH time, $\tau T$, during which the relay harvests the power signal broadcast by the source node, where $0 \leq \tau \leq 1$ is the EH time factor. The remaining time is divided into two equal time slots used for source-to-relay and relay-to-destination information transmission.

Therefore, using (3), the harvested energy at the relay for this system can be written as

$$E_H = \frac{\eta \tau T P_s h_1^2}{d_1^m}, \quad (4)$$

where $0 < \eta < 1$ is the EH efficiency determined mainly by the circuitry. Now, the received signal at the destination node can be expressed as

$$y_d(t) = \sqrt{\frac{P_r}{d_2^m}} h_2 \bar{s}(t) + n_d(t), \quad (5)$$

where $\bar{s}(t)$ is the decoded version of the source signal, $n_d(t) = n_a(t) + n_c(t)$ is the overall noise at the destination node with variance $\sigma_d^2$, $n_c(t)$ is the noise added by the information receiver, and $P_r$ is the relay transmit power which is related to the harvested energy as

$$P_r = \frac{E_H}{(1-\tau)T/2} = \frac{2\eta P_s h_1^2 \tau}{(1-\tau) d_1^m}. \quad (6)$$

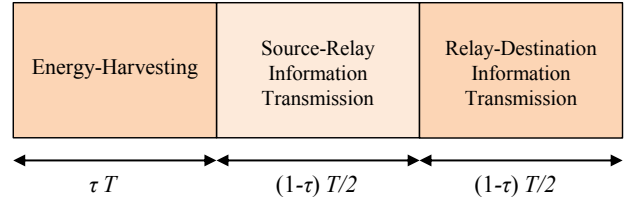

Figure 2: Time frame structure in the TSR protocol.

Substituting (6) into (5) yields

$$y_d(t) = \sqrt{\frac{2\eta\tau P_s}{(1-\tau) d_1^m d_2^m}} h_1 h_2 \bar{s}(t) + n_d(t). \quad (7)$$

Grouping the information and noise terms in (3) and (7), we obtain the SNRs at the relay and destination nodes, respectively, as follows

$$\gamma_r = \frac{P_s h_1^2}{d_1^m \sigma_r^2}, \quad (8)$$

$$\gamma_d = \frac{2\eta\tau P_s h_1^2 h_2^2}{(1-\tau) d_1^m d_2^m \sigma_d^2}. \quad (9)$$

Since in the TSR protocol information transmission takes place only during the time fraction $(1-\tau)$, the instantaneous capacity of the first and second links can be given by

$$C_i^{HD-TSR} = \frac{(1-\tau)}{2} \log_2(1 + \gamma_i) \quad (10)$$

where $i \in \{r, d\}$ and the factor $\frac{1}{2}$ is a result of HD relaying.

To derive the ergodic outage probability for the HD-DF-TSR system, we first write (2) as

$$\begin{aligned}
\mathcal{O}^{TSR}(C_{th}) &= \Pr\{\min\{C_r^{TSR}, C_d^{TSR}\} < C_{th}\} \\
&= 1 - \Pr\{\min\{C_r^{TSR}, C_d^{TSR}\} \geq C_{th}\} \\
&= 1 - \Pr\{C_r^{TSR} \geq C_{th}, C_d^{TSR} \geq C_{th}\} \\
&= 1 - \underbrace{\Pr\{C_r^{TSR} \geq C_{th}\}}_{\mathcal{O}_1^{TSR}(C_{th})} \\
&\quad + \underbrace{\Pr\{C_r^{TSR} \geq C_{th}, C_d^{TSR} < C_{th}\}}_{\mathcal{O}_2^{TSR}(C_{th})}.
\end{aligned} \quad (11)$$

It is clear that the ergodic outage probability requires calculating two probabilities. Using (8) and (10), and substituting $X = h_1^2$, the first probability in (11) can be calculated as

$$\begin{aligned}\mathcal{O}_1^{TSR}(C_{th}) &= \Pr\{C_r^{TSR} \geq C_{th}\} \\ &= \Pr\left\{\frac{(1-\tau)}{2}\log_2\left(1 + \frac{P_s X}{d_1^m \sigma_r^2}\right) \geq C_{th}\right\} \\ &= \Pr\left\{\frac{P_s X}{d_1^m \sigma_r^2} \geq v\right\} \\ &= \Pr\{X \geq a_1 v\} \\ &= 1 - F_X(a_1 v),\end{aligned} \qquad (12)$$

where $v = 2^{\frac{2C_{th}}{1-\tau}} - 1$, $a_1 = d_1^m \sigma_r^2 / P_s$ and $F_X(\cdot)$ denotes the CDF of the RV $X$. Since $X$ is log-normally distributed, its CDF is given by

$$F_X(a_1 v) = 1 - Q\left(\frac{\xi \ln(a_1 v) - 2\mu_{h_2}}{2\sigma_{h_2}}\right), \qquad (13)$$

where $\xi = 10/\ln(10)$ is a scaling constant and $Q(\cdot)$ is the Gaussian Q-function, given by

$$Q(x) = \int_x^\infty \frac{1}{\sqrt{2\pi}} \exp\left(-\frac{t^2}{2}\right) dt. \qquad (14)$$

Now, using (8)–(10), and substituting $Y = h_2^2$, the second probability in (11) can be determined as

$$\begin{aligned}\mathcal{O}_2^{TSR}(C_{th}) &= \Pr\{C_r^{TSR} \geq C_{th}, C_d^{TSR} < C_{th}\} \\ &= \Pr\left\{X \geq a_1 v, \frac{2\eta\tau P_s X Y}{(1-\tau) d_1^m d_2^m \sigma_d^2} < v\right\} \\ &= \Pr\left\{X \geq a_1 v, Y < \frac{a_2 v}{X}\right\},\end{aligned} \qquad (15)$$

where $a_2 = (1-\tau) d_1^m d_2^m \sigma_d^2 / 2\eta\tau P_s$.

Using the PDF and CDF of the log-normally distributed RVs $X$ and $Y$, we can calculate the second probability in (15) as

$$\mathcal{O}_2^{TSR}(C_{th}) = \int_{a_1 v}^\infty f_X(z) F_Y\left(\frac{a_2 v}{z}\right) dz, \qquad (16)$$

where

$$f_X(z) = \frac{\xi}{z\sqrt{8\pi\sigma_{h_1}^2}} \exp\left(-\frac{(\xi \ln(z) - 2\mu_{h_1})^2}{8\sigma_{h_1}^2}\right) \qquad (17)$$

and

$$F_Y\left(\frac{a_2 v}{z}\right) = 1 - Q\left(\frac{\xi \ln\left(\frac{a_2 v}{z}\right) - 2\mu_{h_2}}{2\sigma_{h_2}}\right). \qquad (18)$$

Finally, substituting (12) and (16) into (11) yields the ergodic outage probability of the HD-DF-TSR system, given by

$$\begin{aligned}\mathcal{O}_{TSR}(C_{th}) = &1 - Q\left(\frac{\xi \ln(a_1 v) - 2\mu_{h_2}}{2\sigma_{h_2}}\right) + \frac{\xi}{\sqrt{8\pi\sigma_{h_1}^2}} \\ &\times \int_{a_1 v}^\infty \frac{1}{z} \exp\left(-\frac{(\xi \ln(z) - 2\mu_{h_1})^2}{8\sigma_{h_1}^2}\right) \\ &\times \left(1 - Q\left(\frac{\xi \ln\left(\frac{a_2 v}{z}\right) - 2\mu_{h_2}}{2\sigma_{h_2}}\right)\right) dz.\end{aligned} \qquad (19)$$

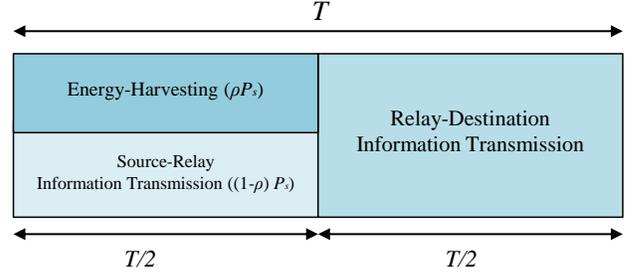

Figure 3: Time frame structure of the PSR protocol.

*2) HD-DF-PSR System:* In the PSR protocol, the block time, $T$, is divided evenly for the source-to-relay and relay-to-destination transmissions as illustrated in Fig. 3. During the first half, the relay allocates a portion of the received signal power, $\rho P$, to the energy-harvester whereas the remaining power, $(1-\rho) P$, is used for information transmission, where $0 \leq \rho \leq 1$ is the power-splitting factor. Therefore, in the first time slot the received signal at the input of the energy-harvester is expressed as

$$\sqrt{\rho} y_r(t) = \sqrt{\frac{\rho P_s}{d_1^m}} h_1 s(t) + \sqrt{\rho} n_a(t). \qquad (20)$$

Using (20), the harvested energy at the relay node can be simply written as

$$E_H = \frac{\eta \rho P_s h_1^2 T}{2 d_1^m}. \qquad (21)$$

On the other hand, the base-band signal at the information receiver, $\sqrt{1-\rho} y_r(t)$, is given by

$$\sqrt{1-\rho} y_r(t) = \sqrt{\frac{(1-\rho) P_s}{d_1^m}} h_2 s(t) + n_r(t), \qquad (22)$$

where $n_r(t) = \sqrt{1-\rho} n_a(t) + n_c(t)$ is the overall noise at the relay with variance $\sigma_r^2 = \sqrt{1-\rho} \sigma_a^2 + \sigma_c^2$.

In the second time slot, the relay decodes the signal in (22), re-modulates and forwards it using the harvested energy in (21). Therefore, the received signal at the destination node can be expressed as

$$y_d(t) = \sqrt{\frac{P_r}{d_2^m}} h_2 \bar{s}(t) + n_d(t), \qquad (23)$$

where $P_r$ is the relay transmit power which is related to the harvested energy as

$$P_r = \frac{E_H}{T/2} = \frac{\eta \rho P_s h_1^2}{d_1^m}. \tag{24}$$

Now, substituting (24) into (23) produces

$$y_d(t) = \sqrt{\frac{\eta \rho P_s}{d_1^m d_2^m}} \, h_1 h_2 \, \bar{s}(t) + n_d(t). \tag{25}$$

Using (22) and (25), the SNRs at the relay and destination nodes can respectively be expressed as

$$\gamma_r = \frac{(1-\rho) P_s h_1^2}{d_1^m \sigma_r^2}, \tag{26}$$

$$\gamma_d = \frac{\eta \rho P_s \, h_1^2 h_2^2}{d_1^m d_2^m \sigma_d^2}. \tag{27}$$

The instantaneous capacity at the relay and destination for the HD-DF-PSR system can be determined using

$$C_i^{PSR} = \frac{1}{2} \log_2(1 + \gamma_i), \tag{28}$$

where $i \in \{r, d\}$.

Similar to the HD-DF-TSR system, the ergodic outage probability of the HD-DF-PSR approach can be calculated as

$$\mathcal{O}^{PSR}(C_{th}) = 1 - \underbrace{\Pr\{C_r^{PSR} \geq C_{th}\}}_{\mathcal{O}_1^{PSR}(C_{th})} + \underbrace{\Pr\{C_r^{PSR} \geq C_{th}, C_d^{PSR} < C_{th}\}}_{\mathcal{O}_2^{PSR}(C_{th})}. \tag{29}$$

Using (26) and (28), the first probability in (29) can be written as

$$\begin{aligned}\mathcal{O}_1^{PSR}(C_{th}) &= \Pr\{C_r^{PSR} \geq C_{th}\} \\ &= \Pr\left\{\frac{1}{2}\log_2\left(1 + \frac{(1-\rho)P_s X}{d_1^m \sigma_r^2}\right) \geq C_{th}\right\} \\ &= \Pr\left\{\frac{(1-\rho)P_s X}{d_1^m \sigma_r^2} \geq R\right\} \\ &= \Pr\{X \geq b_1 R\} \\ &= 1 - F_X(b_1 R),\end{aligned} \tag{30}$$

where $R = 2^{2C_{th}} - 1$, $b_1 = d_1^m \sigma_r^2/(1-\rho)P_s$ and $F_X(\cdot)$ represents the CDF of $X$ which is given in this case as

$$F_Y(b_1 R) = 1 - Q\left(\frac{\xi \ln(b_1 R) - 2\mu_{h_2}}{2\sigma_{h_2}}\right). \tag{31}$$

Using (26)–(28), the second probability in (29) can be calculated as follows

$$\begin{aligned}\mathcal{O}_2^{PSR}(C_{th}) &= \Pr\{C_r^{PSR} \geq C_{th}, C_d^{PSR} < C_{th}\} \\ &= \Pr\left\{X \geq b_1 R, \frac{\eta \rho P_s XY}{d_1^m d_2^m \sigma_d^2} < R\right\} \\ &= \Pr\left\{X \geq b_1 R, Y < \frac{b_2 R}{X}\right\},\end{aligned} \tag{32}$$

where $b_2 = d_1^m d_2^m \sigma_d^2 / \eta \rho P_s$.

Now, using the PDF and CDF of the RVs $X$ and $Y$, we can express $\mathcal{O}_2^{PSR}(C_{th})$ as

$$\mathcal{O}_2^{PSR}(C_{th}) = \int_{b_1 R}^{\infty} f_X(z) F_Y\left(\frac{b_2 R}{z}\right) dz, \tag{33}$$

where

$$f_X(z) = \frac{\xi}{z\sqrt{8\pi\sigma_{h_1}^2}} \exp\left(-\frac{(\xi \ln(z) - 2\mu_{h_1})^2}{8\sigma_{h_1}^2}\right) \tag{34}$$

and

$$F_Y\left(\frac{b_2 R}{z}\right) = 1 - Q\left(\frac{\xi \ln\left(\frac{b_2 R}{z}\right) - 2\mu_{h_2}}{2\sigma_{h_2}}\right). \tag{35}$$

Finally, substituting (30) and (33) into (29) produces the ergodic outage probability of the HD-DF-PSR system in log-normal fading, expressed as

$$\begin{aligned}\mathcal{O}^{PSR}(C_{th}) =\;& 1 - Q\left(\frac{\xi \ln(b_1 R) - 2\mu_{h_2}}{2\sigma_{h_2}}\right) + \frac{\xi}{\sqrt{8\pi\sigma_{h_1}^2}} \\ & \times \int_{b_1 R}^{\infty} \frac{1}{z} \exp\left(-\frac{(\xi \ln(z) - 2\mu_{h_1})^2}{8\sigma_{h_1}^2}\right) \\ & \times \left(1 - Q\left(\frac{\xi \ln\left(\frac{b_2 R}{z}\right) - 2\mu_{h_2}}{2\sigma_{h_2}}\right)\right) dz.\end{aligned} \tag{36}$$

*3) HD-DF-IRR System:* Unlike the TSR and PSR protocols, the IRR scheme is capable of concurrently processing information and harvesting energy from the same received signal. Therefore, the signal received at the information receiver of the relay can be expressed as

$$y_r(t) = \sqrt{\frac{P_s}{d_1^m}} \, h_1 s(t) + n_r(t), \tag{37}$$

and the harvested energy and the relay transmit power can therefore be given respectively as

$$E_H = \frac{\eta P_s h_1^2}{d_1^m} (T/2), \tag{38}$$

$$P_r = \frac{2E_H}{T} = \frac{\eta P_s h_1^2}{d_1^m}. \tag{39}$$

Using (39), the received signal at the destination can be written as

$$y_d(t) = \sqrt{\frac{\eta P_s}{d_1^m d_2^m}} h_1 h_2 \bar{s}(t) + n_d(t). \quad (40)$$

Now, using (37) and (40), the SNRs at the relay and destination nodes in the HD-DF-IRR system can be respectively expressed as

$$\gamma_r = \frac{P_s h_1^2}{d_1^m \sigma_r^2} \quad (41)$$

and

$$\gamma_d = \frac{\eta P_s h_1^2 h_2^2}{d_1^m d_2^m \sigma_d^2}. \quad (42)$$

The ergodic outage probability of the HD-DF-TSR system can be calculated as

$$\mathcal{O}^{IRR}(C_{th}) = 1 - \underbrace{\Pr\{C_r^{IRR} \geq C_{th}\}}_{\mathcal{O}_1^{IRR}(C_{th})} + \underbrace{\Pr\{C_r^{IRR} \geq C_{th}, C_d^{IRR} < C_{th}\}}_{\mathcal{O}_2^{IRR}(C_{th})}. \quad (43)$$

The first probability $\mathcal{O}_1^{IRR}(C_{th})$ can be given by

$$\begin{aligned}\mathcal{O}_1^{IRR}(C_{th}) &= \Pr\{C_r^{IRR} \geq C_{th}\} \\ &= \Pr\left\{\frac{1}{2}\log_2\left(1 + \frac{P_s X}{d_1^m \sigma_r^2}\right) \geq C_{th}\right\} \\ &= \Pr\left\{\frac{P_s X}{d_1^m \sigma_r^2} \geq R\right\} \\ &= \Pr\{X \geq c_1 R\} \\ &= 1 - F_X(c_1 R),\end{aligned} \quad (44)$$

where $c_1 = d_1^m \sigma_r^2 / P_s$ and $F_X(\cdot)$ represents the CDF of $X$ expressed in this case as

$$F_Y(c_1 R) = 1 - Q\left(\frac{\xi \ln(c_1 R) - 2\mu_{h_2}}{2\sigma_{h_2}}\right). \quad (45)$$

Using (28), (41) and (42), the second probability in (43) can be written as

$$\begin{aligned}\mathcal{O}_2^{IRR}(C_{th}) &= \Pr\{C_r^{IRR} \geq C_{th}, C_d^{IRR} < C_{th}\} \\ &= \Pr\left\{X \geq c_1 R, \frac{\eta P_s XY}{d_1^m d_2^m \sigma_d^2} < R\right\} \\ &= \Pr\left\{X \geq c_1 R, Y < \frac{c_2 R}{X}\right\},\end{aligned} \quad (46)$$

where $c_2 = d_1^m d_2^m \sigma_d^2 / \eta P_s$.

Using the PDF and CDF of the RVs $X$ and $Y$, we get

$$\mathcal{O}_2^{IRR}(C_{th}) = \int_{c_1 R}^{\infty} f_X(z) F_Y\left(\frac{c_2 R}{z}\right) dz, \quad (47)$$

where

$$f_X(z) = \frac{\xi}{z\sqrt{8\pi\sigma_{h_1}^2}} \exp\left(-\frac{(\xi \ln(z) - 2\mu_{h_1})^2}{8\sigma_{h_1}^2}\right) \quad (48)$$

and

$$F_Y\left(\frac{c_2 R}{z}\right) = 1 - Q\left(\frac{\xi \ln\left(\frac{c_2 R}{z}\right) - 2\mu_{h_2}}{2\sigma_{h_2}}\right). \quad (49)$$

Finally, substituting (44) and (47) into (43) we obtain the ergodic outage probability of the HD-DF-IRR system as

$$\begin{aligned}\mathcal{O}^{IRR}(C_{th}) = 1 &- Q\left(\frac{\xi \ln(c_1 R) - 2\mu_{h_2}}{2\sigma_{h_2}}\right) + \frac{\xi}{\sqrt{8\pi\sigma_{h_1}^2}} \\ &\times \int_{c_1 R}^{\infty} \frac{1}{z} \exp\left(-\frac{(\xi \ln(z) - 2\mu_{h_1})^2}{8\sigma_{h_1}^2}\right) \\ &\times \left(1 - Q\left(\frac{\xi \ln\left(\frac{c_2 R}{z}\right) - 2\mu_{h_2}}{2\sigma_{h_2}}\right)\right) dz.\end{aligned} \quad (50)$$

### B. Half-Duplex with AF Relaying

In this section, we analyze the ergodic outage probability of the HD system with AF relaying for different EH protocols. Although some of the results below have been derived in [29], we feel that presenting them here is important to make the current work self-contained.

*1) HD-AF-TSR System:* In this system, the signal received at the relay during the EH time is given by (3), and therefore the harvested energy is equal to (4). After the baseband processing and amplification at the relay, the relay transmit signal can be written as

$$r(t) = \sqrt{\frac{P_s P_r}{d_1^m}} G h_1 s(t) + \sqrt{P_r} G n_r(t), \quad (51)$$

where $P_r$ is given by (6), $n_r(t) = n_a(t) + n_c(t)$ with variance $\sigma_r^2 = \sigma_a^2 + \sigma_c^2$ and $G$ is the relay gain given by $G = \frac{1}{\sqrt{\frac{P_s}{d_1^m} h_1^2 + \sigma_r^2}}$. The received signal at the destination can then be expressed as

$$y_d(t) = \sqrt{\frac{P_s P_r}{d_1^m d_2^m}} G h_1 h_2 s(t) + \sqrt{\frac{P_r}{d_2^m}} G h_2 n_r(t) + n_d(t). \quad (52)$$

Using (6) and (52), along with some basic algebraic manipulations, the SNR at the destination can be obtained as

$$\gamma_d = \frac{2\eta\tau P_s h_1^2 h_2^2}{2\eta\tau d_1^m \sigma_r^2 h_2^2 + (1-\tau) d_1^m d_2^m \sigma_d^2}, \quad (53)$$

which can also be simplified to

$$\gamma_d = \frac{a_1 h_1^2 h_2^2}{a_3 h_2^2 + a_2}, \quad (54)$$

where $a_1 = 2\eta\tau P_s$, $a_2 = 2\eta\tau d_1^m \sigma_r^2$ and $a_3 = (1-\tau) d_1^m d_2^m \sigma_d^2$.

Now, substituting (54) into (1), we can determine the ergodic outage probability for the HD-AF-TSR system as

$$\mathcal{O}^{TSR}(C_{th}) = \Pr\left\{\frac{a_1 X Y}{a_3 Y + a_2} < v\right\}$$
$$= \Pr\left\{Y < \frac{a_2 v}{a_1 X - a_3 v}\right\}. \quad (55)$$

Since $Y$ is always a positive value, we can express the probability above as

$$\mathcal{O}^{TSR}(C_{th}) = \begin{cases} \Pr\left(Y < \frac{v a_2}{a_1 X - v a_3}\right), & X < \frac{v a_3}{a_1} \\ \Pr\left(Y > \frac{v a_2}{a_1 X - v a_3}\right) = 1, & X > \frac{v a_3}{a_1}. \end{cases} \quad (56)$$

The ergodic outage probability can now be calculated as

$$\mathcal{O}^{TSR}(C_{th}) = \int_0^{\frac{v a_3}{a_1}} f_X(z)\, dz + \int_{\frac{v a_3}{a_1}}^{\infty} f_X(z) \underbrace{\Pr\left(Y \leq \frac{v a_2}{a_1 z - v a_3}\right)}_{F_Y(\cdot)} dz \quad (57)$$

where $f_X(\cdot)$ is the PDF of $X$ and the probability in the second integral represents the CDF of $Y$, i.e., $F_Y(\cdot)$; these PDF and CDF can be given respectively as

$$f_X(z) = \frac{\xi}{z\sqrt{8\pi\sigma_{h_1}^2}} \exp\left(-\frac{(\xi \ln(z) - 2\mu_{h_1} - \xi \ln(a_1))^2}{8\sigma_{h_1}^2}\right) \quad (58)$$

and

$$F_Y(\Gamma) = 1 - Q\left(\frac{\xi \ln(\Gamma) - 2\mu_{h_2}}{2\sigma_{h_2}}\right), \quad (59)$$

where $\Gamma = \frac{v a_3}{z - v a_2}$.

Finally, substituting (58) and (59) into (57) gives the ergodic outage probability of the HF-AF-TSR system as

$$\mathcal{O}^{TSR}(C_{th}) = 1 - \frac{\xi}{\sqrt{8\pi\sigma_{h_1}^2}} \int_{v a_2}^{\infty} \frac{1}{z} Q\left(\frac{\xi \ln(\Gamma) - 2\mu_{h_2}}{2\sigma_{h_2}}\right)$$
$$\times \exp\left(-\frac{(\xi \ln(z) - (2\mu_{h_1} + \xi \ln(a_1)))^2}{8\sigma_{h_1}^2}\right) dz. \quad (60)$$

*2) HD-AF-PSR System:* In this system, the received signal at the input of the energy-harvester can also be given by (20), and therefore the harvested energy is equal to (21). In light of this, the relay transmit signal can be expressed as

$$r(t) = \sqrt{\frac{(1-\rho) P_s P_r}{d_1^m}} G h_1 s(t) + \sqrt{P_r} G n_r(t), \quad (61)$$

where $n_r(t) = \sqrt{1-\rho}\, n_a(t) + n_c(t)$, $P_r$ is defined in (24) and $G$ is given for the HD-AF-PSR system as $G = \frac{1}{\sqrt{\frac{(1-\rho)P_s}{d_1^m} h_1^2 + \sigma_r^2}}$. We can now express the signal received at the destination node as

$$y_d(t) = \sqrt{\frac{(1-\rho) P_s P_r}{d_1^m d_2^m}} G h_1 h_2 s(t) + \sqrt{\frac{P_r}{d_2^m}} G h_2 n_r(t) + n_d(t). \quad (62)$$

Substituting (24) into (62) and with some basic algebraic manipulations, we obtain the SNR at the destination as

$$\gamma_d = \frac{\eta\rho(1-\rho) P_s h_1^2 h_2^2}{\eta\rho d_1^m \sigma_c^2 h_2^2 + \eta\rho(1-\rho) d_1^m \sigma_a^2 h_2^2 + (1-\rho) d_1^m d_2^m \sigma_d^2}. \quad (63)$$

Using $b_1 = \eta\rho(1-\rho) P_s$, $b_2 = \eta\rho d_1^m \sigma_c^2$, $b_3 = \eta\rho(1-\rho) d_1^m \sigma_a^2$ and $b_4 = (1-\rho) d_1^m d_2^m \sigma_d^2$, we can rewrite (63) in the following simplified form

$$\gamma_d = \frac{b_1 h_1^2 h_2^2}{b_2 h_2^2 + b_3 h_2^2 + b_4}. \quad (64)$$

Following the same procedure as in the HD-AF-TSR system, it is straightforward to show that the ergodic outage probability of the HD-AF-PSR system can be calculated as

$$\mathcal{O}^{PSR}(C_{th}) = 1 - \frac{\xi}{\sqrt{8\pi\sigma_{h_1}^2}} \int_{v(b_2+b_3)}^{\infty} \frac{1}{z} Q\left(\frac{\xi \ln(\Lambda) - 2\mu_{h_2}}{2\sigma_{h_2}}\right)$$
$$\times \exp\left(-\frac{(\xi \ln(z) - 2\mu_{h_1} - \xi \ln(b_1))^2}{8\sigma_{h_1}^2}\right) dz, \quad (65)$$

where $\Lambda = \frac{v b_4}{z - v b_2 - v b_3}$. For more details the reader may refer to [29].

*3) HD-AF-IRR System:* The harvested energy and relay transmit power in this system can also be given by (38) and (39), respectively. Recalling that the relay has a gain of $G$, we can write the received signal at the destination as

$$y_d(t) = \sqrt{\frac{P_s P_r}{d_1^m d_2^m}} G h_1 h_2 s(t) + \sqrt{\frac{P_r}{d_2^m}} G h_2 n_r(t) + n_d(t), \quad (66)$$

where $G = \frac{1}{\sqrt{\frac{P_s}{d_1^m} h_1^2 + \sigma_r^2}}$.

Substituting (39) into (66), and then grouping the information and noise signals, the SNR at the destination node can be given by

$$\gamma_d = \frac{\eta P_s h_1^2 h_2^2}{\eta d_1^m \sigma_r^2 h_2^2 + d_1^m d_2^m \sigma_d^2}, \quad (67)$$

which can also be written, for more convenience, as

$$\gamma_d = \frac{c_1 h_1^2 h_2^2}{c_2 h_2^2 + c_3}, \quad (68)$$

where $c_1 = \eta P_s$, $c_2 = \eta d_1^m \sigma_r^2$ and $c_3 = d_1^m d_2^m \sigma_d^2$.

Following the same procedure as in Sec. III-B-3, we can show that the ergodic outage probability of the HD-AF-IRR system can be calculated as

$$\mathcal{O}^{IRR}(C_{th}) = 1 - \frac{\xi}{\sqrt{8\pi\sigma_{h_1}^2}} \int_{vc_2}^{\infty} \frac{1}{z} Q\left(\frac{\xi \ln(\Upsilon) - 2\mu_{h_2}}{2\sigma_{h_2}}\right)$$
$$\times \exp\left(-\frac{(\xi \ln(z) - 2\mu_{h_1} - \xi \ln(c_1))^2}{8\sigma_{h_1}^2}\right) dz, \quad (69)$$

where $\Upsilon = \frac{vc_3}{z - vc_2}$.

## IV. FULL-DUPLEX RELAYING SYSTEM

This section is dedicated to analyze the performance of a dual-hop FD network with DF and AF relaying based on the TSR EH protocol. Recall that unlike HD, in FD the relay is equipped with two antennas and this allows simultaneous reception and transmission of information at the relay, see Fig. 1(b).

To begin with, in the first time slot, the received signal at the relay can also be given by (3) and the relay transmit power will have the following form

$$P_r = \frac{E_H}{(1-\tau)T} = \frac{\eta \tau P_s h_1^2}{(1-\tau) d_1^m}. \quad (70)$$

In the second time slot (information transmission phase), the received signal at the relay is given by

$$y_r(t) = \sqrt{\frac{P_s}{d_1^m}} h_1 s(t) + g r(t) + n_a(t), \quad (71)$$

where $g$ is the loop-back interference channel and $r(t)$ is the loop-back interference signal due to FD relaying and it satisfies the following $E\left\{|r(t)|^2\right\} = P_r$.

It should be pointed out that in the FD system, since the relay knows its own signal, it usually applies interference cancellation to reduce the loop-back interference. Therefore, we can now write the post-cancellation signal at the relay as

$$y_r(t) = \sqrt{\frac{P_s}{d_1^m}} h_1 s(t) + \hat{g} \hat{r}(t) + n_a(t), \quad (72)$$

where $\hat{g}$ is the residual loop-back interference channel caused by imperfect interference cancellation and $E\left\{|\hat{r}(t)|^2\right\} = P_r$. Below, we derive the ergodic outage probability expressions for both FD-DF and FD-AF relaying systems.

### A. Full-Duplex with DF Relaying (FD-DF-TSR)

When DF relaying is applied, the relay will decode and forward the source signal; hence, the received signal at the destination can be expressed as

$$y_d(t) = \sqrt{\frac{P_r}{d_2^m}} h_2 \bar{s}(t) + n_d(t). \quad (73)$$

Using (70), (72) and (73), the SNRs at the relay and destination in the FD-DF-TSR system can be written respectively as

$$\gamma_r = \frac{P_s h_1^2}{P_r d_1^m g^2}$$
$$= \frac{1-\tau}{\eta \tau g^2}, \quad (74)$$

and

$$\gamma_d = \frac{P_r h_2^2}{d_2^m \sigma_d^2}$$
$$= \frac{\eta \tau P_s h_1^2 h_2^2}{(1-\tau) d_1^m d_2^m \sigma_d^2}. \quad (75)$$

The instantaneous capacity of the source-to-relay and relay-to-destination links can now be expressed as

$$C_i^{FD-DF} = (1-\tau) \log_2(1+\gamma_i), \quad (76)$$

where $i \in \{r, d\}$. Comparing (10) and (76), it is obvious that the factor $\frac{1}{2}$ is no longer present in (76) due to the FD nature of the relay.

Now, the ergodic outage probability for this system can be determined as

$$\mathcal{O}^{FD-DF}(C_{th}) = \Pr\left\{\min\left\{C_r^{FD-DF}, C_d^{FD-DF}\right\} < C_{th}\right\}. \quad (77)$$

Using (74)-(76), and substituting $Z = h_1^2 h_2^2$ and $W = g^2$, the probability in (77) can be expressed as

$$\mathcal{O}^{FD-DF}(C_{th}) = \Pr\left\{\min\left\{\frac{1-\tau}{\eta \tau W}, \frac{\eta \tau P_s Z}{(1-\tau) d_1^m d_2^m \sigma_d^2}\right\} < v\right\}, \quad (78)$$

where $v = 2^{\frac{C_{th}}{1-\tau}} - 1$.

Using the fact that the RVs $Z$ and $W$ are independent, the ergodic outage probability of the FD-DF-TSR system can be given by

$$\mathcal{O}^{FD-DF}(C_{th}) = 1 - \bar{F}_W\left(\frac{\eta\tau}{1-\tau}v\right)\bar{F}_Z\left(\frac{(1-\tau)d_1^m d_2^m \sigma_d^2}{\eta\tau P_s}v\right) \quad (79)$$

where $\bar{F}_W(\cdot)$ and $\bar{F}_Z(\cdot)$ are the CCDFs of $W$ and $Z$, respectively.

The first CCDF, $\bar{F}_W(\cdot)$, is straightforward to obtain since the RV $W$ is log-normally distributed. On the other hand, the RV $W$ is a product of two log-normally distributed RVs. Using the properties of the log-normal distribution, it can be shown that the CCDF of $Z$ is

$$\bar{F}_Z(\cdot) = Q\left(\frac{\xi \ln(\Delta) - 2(\mu_{h_1} + \mu_{h_2})}{\sqrt{2}(\sigma_{h_1} + \sigma_{h_2})}\right), \quad (80)$$

where $\Delta = \frac{(1-\tau)d_1^m d_2^m \sigma_d^2}{\eta\tau P_s}v$.

With this in mind, we can finally express the ergodic outage probability of the FD-DF-TSR system as

$$\mathcal{O}^{FD-DF}(C_{th}) = 1 - Q\left(\frac{\xi \ln\left(\frac{\eta\tau}{1-\tau}v\right) + 2\mu_g}{2\sigma_g}\right)$$

$$\times\ Q\left(\frac{\xi \ln(\Delta) - 2\sum_{i\in\{1,2\}} \mu_{h_i}}{\sqrt{2}\sum_{i\in\{1,2\}} \sigma_{h_i}}\right). \quad (81)$$

### B. Full-Duplex with AF Relaying (FD-AF-TSR)

In the case of FD-AF system, the received signal at the relay (72) will be amplified and then forwarded to the destination node. Hence, the received signal at the destination can be expressed as

$$y_r(t) = \sqrt{\frac{P_s P_r}{d_1^m d_2^m}} h_1 h_2 G\, s(t) + \sqrt{\frac{P_r}{d_2^m}} h_2 g G\, r(t)$$

$$+ \sqrt{\frac{P_r}{d_2^m}} h_2 G n_a(t) + n_d(t), \quad (82)$$

where $P_r$ is given by (70) and $G$ is the relay gain defined as

$$G = \frac{1}{\sqrt{\frac{P_s}{d_1^m}h_1^2 + \hat{g}^2 P_r + \sigma_r^2}}. \quad (83)$$

Grouping the information signal and noise terms in (82), we can write the SNR at the destination as

$$\gamma_d = \frac{P_s h_1^2 h_2^2}{d_1^m d_2^m g^2 \left(\frac{P_s h_1^2 \sigma_r^2}{P_r g^2 d_1^m} + \frac{P_r h_2^2}{d_2^m} + \sigma_r^2\right)}. \quad (84)$$

Note that the instantaneous capacity of the FD-AF-TSR system is determined using

$$C_d^{FD-AF} = (1-\tau)\log_2(1+\gamma_d). \quad (85)$$

Similar to the DF scenario, comparing (10) and (85), it can evident that the factor $\frac{1}{2}$ is no longer present (85) due to the FD relaying.

Now, using the definition in (1) along with (84) and (85), we can calculate the ergodic outage probability of the FD-AF-TSR system as

$$\mathcal{O}^{FD-AF}(C_{th}) = \Pr\left\{C_d^{FD-AF}(\gamma_d) < C_{th}\right\},$$

$$= \Pr\left\{\frac{P_s h_1^2 h_2^2 d_1^{-m} d_2^{-m}}{g^2\left(\frac{P_s h_1^2 \sigma_r^2}{P_r g^2 d_1^m} + \frac{P_r h_2^2}{d_2^m} + \sigma_r^2\right)} < v\right\} \quad (86)$$

where $v = 2^{\frac{C_{th}}{1-\tau}} - 1$.

Using the substitutions $k = \eta\tau/(1-\tau)$, $Z = h_1^2 h_2^2$ and $W = g^2$, along with some algebraic manipulations, we can rewrite (86) as

$$\mathcal{O}^{FD-AF}(C_{th}) = \Pr\left\{Z < \frac{d_1^m d_2^m v \sigma_r^2\left(\frac{1}{k} + W\right)}{P_s - P_s k\, v\, W}\right\}. \quad (87)$$

We know that $Z$ is always a positive value; hence, the probability in (87) can be rewritten in the following form

$$\mathcal{O}^{FD-AF}(C_{th}) = \begin{cases} \Pr\left(Z \leq \frac{d_1^m d_2^m v \sigma_r^2\left(\frac{1}{k}+W\right)}{P_s - P_s k\, v\, W}\right), & W < \frac{1}{kv} \\ \Pr\left(Z > \frac{d_1^m d_2^m v \sigma_r^2\left(\frac{1}{k}+W\right)}{P_s - P_s k\, v\, W}\right) = 1, & W > \frac{1}{kv}. \end{cases} \quad (88)$$

This probability can be calculated as follows

$$\mathcal{O}^{FD-AF}(C_{th}) = 1 - \int_0^{\frac{1}{kv}} f_W(z)\, \bar{F}_Z\left(\frac{d_1^m d_2^m v \sigma_r^2\left(\frac{1}{k}+z\right)}{P_s - P_s k\, v\, z}\right) dz, \quad (89)$$

where $f_W(\cdot)$ is the PDF of $W$ given by

$$f_W(z) = \frac{\xi}{z\sqrt{8\pi\sigma_g^2}}\exp\left(-\frac{(\xi\ln(z) - 2\mu_g)^2}{8\sigma_g^2}\right) \quad (90)$$

and $\bar{F}_Z(\cdot)$ is the CCDF of $Z$ defined as

$$\bar{F}_Z(z) = Q\left(\frac{\xi \ln\left(\frac{d_1^m d_2^m v \sigma_r^2\left(\frac{1}{k}+z\right)}{P_s - P_s k\, v\, z}\right) - 2\sum_{i\in\{1,2\}} \mu_{h_i}}{\sqrt{2}\sum_{i\in\{1,2\}} \sigma_{h_i}}\right). \quad (91)$$

Finally, substituting (90) and (91) into (89), we can express the ergodic outage probability of the FD-AF-TSR system as in (92), shown at the top of the next page, where $\Gamma = \frac{d_1^m d_2^m v \sigma_r^2\left(\frac{1}{k}+z\right)}{P_s - P_s k\, v\, z}$.

$$\mathcal{O}^{FD-AF}(C_{th}) = 1 - \frac{\xi}{\sqrt{8\pi\sigma_{h_1}^2}} \int_0^{\frac{1}{kv}} \frac{1}{z} Q\left(\frac{\xi \ln(\Gamma) - 2\sum_{i\in\{1,2\}}\mu_{h_i}}{\sqrt{2}\sum_{i\in\{1,2\}}\sigma_{h_i}}\right) \exp\left(-\frac{(\xi \ln(z) - 2\mu_{h_1})^2}{8\sigma_{h_1}^2}\right) dz \qquad (92)$$

## V. RESULTS AND DISCUSSIONS

In this section, we present some numerical examples of the derived expressions above along with Monte Carlo simulations. Unless specified otherwise, we use in our evaluations the following system parameters: $P_s = 1$W, $\eta = 1$, $m = 2$, $d_1 = d_2 = 5$ m, $C_{th} = 2$ bps/Hz, $\sigma_1^2 = \sigma_2^2 = 4$ dB, $\mu_1 = \mu_2 = 3$ dB and $\sigma_r^2 = \sigma_d^2 = 2\sigma_{ra}^2 = 2\sigma_{rc}^2 = 0.005$W.

### A. The Impact of $\tau$ and $\rho$ on the Performance of HD-DF and HD-AF Systems

This section discusses and compares the effect of the EH parameters $\tau$ (EH time factor in the TAR approach) and $\rho$ (power splitting factor in the PSR approach) on the ergodic outage probability of the HD-DF-TSR, HD-DF-PSR, HD-AF-TSR and HD-AF-PSR systems. To achieve this, we plot in Fig. 4 the analytical and simulated ergodic outage probability for the four systems as a function of $\tau$ and $\rho$. Note that the analytical results are obtained from (19), (36), (60) and (65) for these systems, respectively. The good agreement between the analytical and simulated results clearly indicates the accuracy of our analysis. It can be seen from the results in Fig. 4 that DF relaying, in both TSR and PSR systems, tends to offer slightly better performance in comparison to that of AF relaying. This is mainly because the processing energy cost of DF relaying, which is of course higher than that of AF relaying, is ignored here; this will be investigated in more details later. In addition, it is worthwhile pointing out that this improvement is more pronounced in the TSR-based system compared to that in the PSR-based approach. Another interesting observation one can see from these results is the significant deterioration in the ergodic outage probability when $\tau$ or $\rho$ approach either zero or one due to the fact that the harvested energy becomes either too small or unnecessarily too large (leading to no resources left for data transmission). Therefore, optimizing these parameters is of great importance to maximize the system performance; this phenomena is investigated thoroughly below. It is worth mentioning that it is possible to improve the performance by allocating the channel unequally depending on the relative channel distributions in both TSR and PSR protocols.

### B. Performance Optimization and Impact of Log-normal Fading Parameters

In this section, we address the optimization problem of the EH time factor and the power-splitting factor for the TSR and PSR systems. In addition, we also discuss the impact of log-normal fading, i.e., the impact of the distribution parameters, on the behavior of the different protocols deployed in this work. In this respect, we show in Fig. 5 numerical results of the minimum achievable ergodic outage probability as a function of the source-to-relay and relay-to-destination channel variances for the HD-DF-TSR, HD-DF-PSR, HD-AF-TSR and HD-AF-PSR systems with different transmit power values; these results are obtained using (19), (36), (60) and (65), respectively. Note that although it is very difficult to get the solution in closed-form, it does not pose any difficulty to obtain numerical solutions using software tools. For comparison's sake, results for the HD-DF-IRR and HD-AF-IRR are also included on this figure, which are obtained from (50) and (69), respectively. A number of observations can be seen from this figure. For instance, as opposed to the Rayleigh fading case, it is observed that increasing the channel variance will always degrade performance for all the considered systems. Similar to the previous section, results in Fig. 5 indicate that DF-based systems always outperform the AF-based ones throughout the channel variance spectrum. It is also noticeable that the optimized PST schemes perform better than the optimized TSR systems and that the IRR system serves as a lower bound. Furthermore, comparing Figs. 5(a) and 5(b), it can be noted that increasing the source transmit power considerably enhances the ergodic outage probability for all the system configurations, which is intuitive.

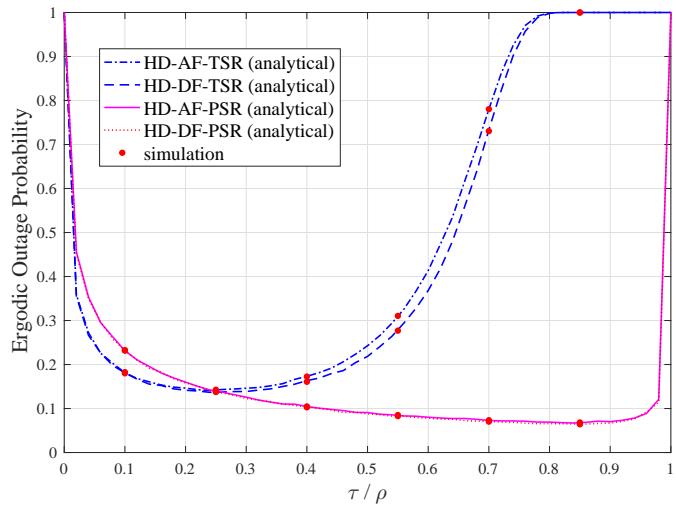

Figure 4: Ergodic outage probability performance with respect to $\tau$ and $\rho$ for the HD-DF-TSR, HD-DF-PSR, HD-AF-TSR and HD-AF-PSR systems.

### C. Impact of Processing and Channel Estimation Energy Cost

Although the assumptions of perfect channel state information availability and the zero processing energy cost at the DF relay have simplified our analysis in this work, such assumptions are unrealistic in practice. We therefore dedicate this section to examine the impact of the energy cost at the DF relay due to information processing and channel estimation; the combined power cost will be referred to as $(P_c)$. The focus here will particularly be kept on the IRR protocol and the end-to-end distance is fixed at 30m. With this in mind, Fig. 6 demonstrates the ergodic outage probability for both the HD-AF-IRR and HD-DF-IRR with respect to the source-to-relay distance with different values of $P_c$; specifically, we consider $P_c = 0$, $0.01P_r$, and $0.02P_r$W, which correspond to $0\%$, $1\%$ and $2\%$ of the harvested power.

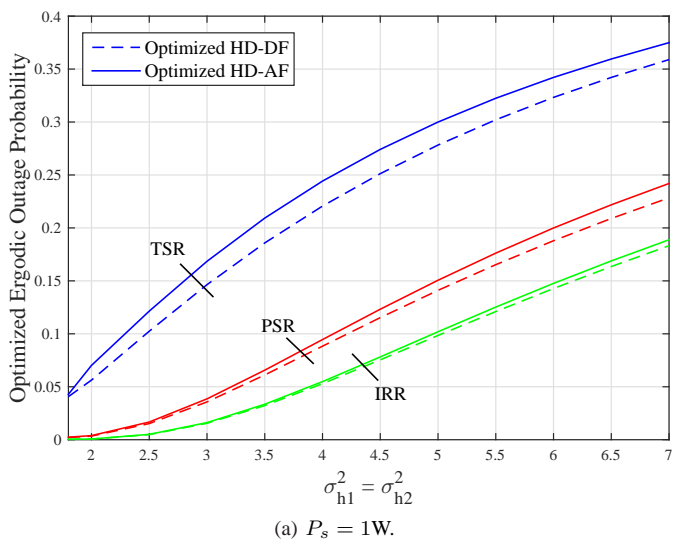

(a) $P_s = 1$W.

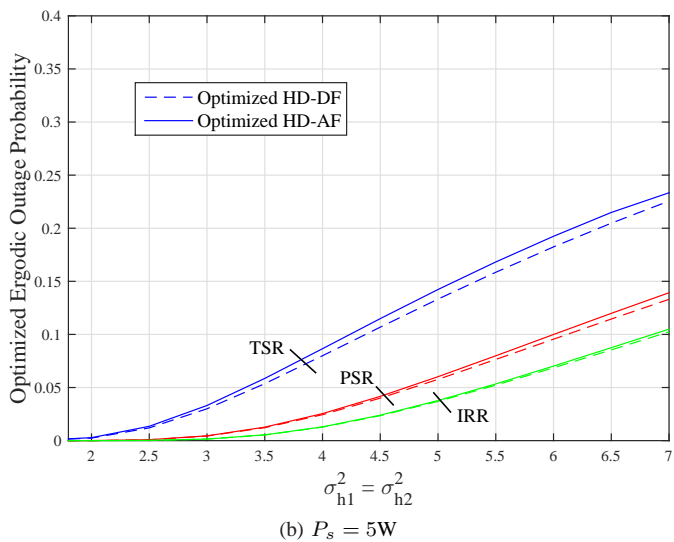

(b) $P_s = 5$W

Figure 5: Minimum achievable ergodic outage probability versus the source-to-relay and relay-to-destination channel variance for the HD-DF-TSR, HD-DF-PSR, HD-DF-IRR, HD-AF-TSR, HD-AF-PSR and HD-AF-IRR systems when $P_s = 1$W and 5W.

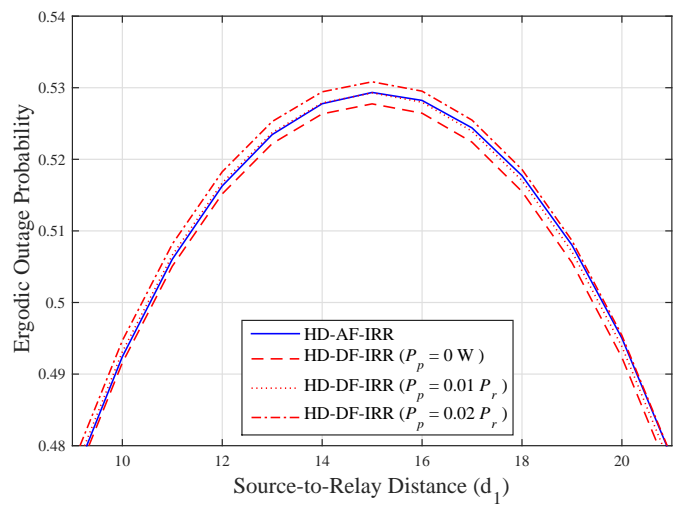

Figure 6: Ergodic outage probability versus the source-to-relay distance for the HD-DF-IRR and HD-AF-IRR systems with different values of $P_c$ at the DF relay when $d_1 + d_2 = 30$m.

As one can readily observe from this figure, the ideal HD-DF-IRR scheme, i.e., $P_c = 0$W, always outperforms the HD-AF-IRR system regardless of the position of the relay. However, when the energy consumption at the DF relay is taken into account, the performance of both DF and AF relaying systems will become very comparable, e.g., $P_c = 0.01P_r$W. However, when $P_c$ increases further, i.e., $P_c = 0.02P_r$W, AF relaying can start to perform better than DF relaying making the former more attractive in some practical scenarios. Another interesting remark on the results in Fig. 6 is that the poorest probability performance, for AF and DF relaying alike, is experienced when the relay is placed at the midpoint between the source and destination nodes. This is simply justified by the fact that at this point the relay will need more time harvesting energy and this consequently impacts the information transmission time; hence, high ergodic outage probability occurs.

### D. FD versus HD Relaying

The performance of the FD and HD systems with DF and AF relaying is discussed in this section. Fig. 7 depicts the ergodic outage probability as a function of the transmission rate threshold for the FD-DF and FD-AF systems with various values of the loop-back interference channel variance and the source transmit power; more specifically, $\sigma_g^2 = \{2, 5\text{dB}\}$ and $P_s = \{1, 10\text{W}\}$. Results for the HD-DF and HD-AF schemes are also included in this figure. Note that the analytical results of the FD-DF and FD-AF systems are obtained from (81) and (92), respectively. The system parameters used in here are: $d_1 = d_2 = 5$m, $\tau = 0.01$, $\sigma_1^2 = \sigma_2^2 = 4$dB and $\mu_1 = \mu_2 = \mu_g = 3$dB. It is can be seen from Fig. 7(a) that when the loop-back interference channel variance is $\sigma_g^2 = 5$dB, the FD schemes always outperform the HD ones for the same transmission rate irrespective of the value of $P_s$. On the other hand, however, it is demonstrated in Fig. 7(b) that when the loop-back interference channel variance $\sigma_g^2 = 2$dB, HD outperforms FD. The final remark on these results is that FD-DF relaying has better performance than FD-AF relaying in all the protocols deployed.

### VI. CONCLUSIONS

This paper studied the ergodic outage probability of HD and FD relaying in EH networks over log-normal fading channels.

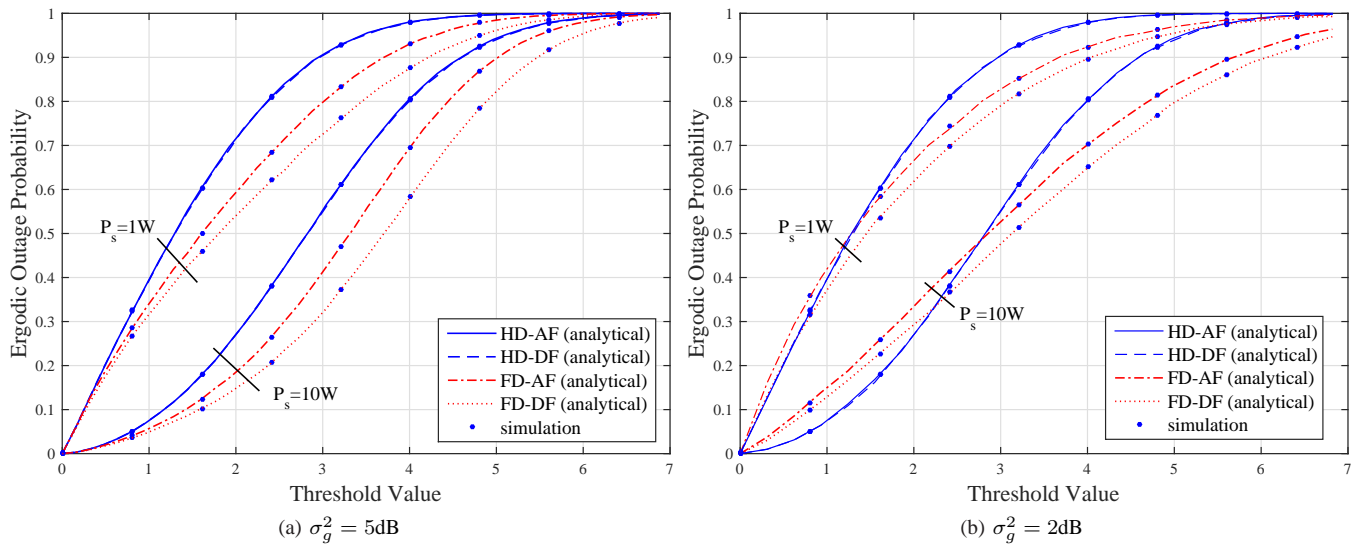

Figure 7: Ergodic outage probability with respect to the transmission rate value for the FD-DF, FD-AF, HD-DF and HD-AF systems with different values of $\sigma_g^2$ and $P_s$. All results in this figure are based on the TSR protocol.

More specifically, we investigated the performance of both AF and DF relaying with three well-known EH protocols, namely, TSR, PSR and IRR. Accurate analytical expressions for the ergodic outage probability for these systems were derived and then validated with computer simulations. It was demonstrated that DF relaying is able to always offer better probability performance compared to AF relaying when the processing energy cost for the former at the relay is ignored. However, when the processing energy cost is taken into account AF relaying may outperform DF relaying. It was also shown that increasing the variance of the log-normal fading channel will degrade the performance. Comparing the performances of FD and HD relaying systems, it was found that FD relaying can considerably enhance the system performance as long as the loop-back interference due FD relaying is relatively low. However, if this interference increases, performance may severely degrade and consequently HD relaying can perform better. Finally, it is to be pointed out that a relay channel with the direct link will be a subject of future research.

ACKNOWLEDGMENT

This research has been carried out within the "Smart In-Building Micro Grid for Energy Management" project funded by EPSRC (EP/M506758/1) and supported by Innovate UK (Innovate UK Project 101836).